\documentclass[11pt,a4paper]{article}
\pdfoutput=1
\usepackage{jheppub}
\usepackage{tikz}
\usetikzlibrary{intersections, calc, arrows.meta, bending}
\usepackage{subcaption}
\usepackage{url}

\DeclareMathOperator{\Tr}{Tr}

\newcommand{\ri}{\mathrm{i}}
\renewcommand{\th}{\theta}

\newcommand{\cob}{\delta}

\newcommand{\vep}{\varepsilon}
\newcommand{\vth}{\vartheta}

\newcommand{\hf}{\frac{1}{2}}

\newcommand{\si}{\sigma}

\newcommand{\del}{\partial}

\newcommand{\lap}{\Delta}
\newcommand{\bra}{\langle}
\newcommand{\ket}{\rangle}
\newcommand{\la}{\lambda}

\newcommand{\h}[1]{\widehat{#1}}
\newcommand{\bt}{\beta}

\newcommand{\rt}[1]{\sqrt{#1}}
\newcommand{\cO}{\mathcal{O}}
\newcommand{\cZ}{\mathcal{Z}}

\newcommand{\cH}{\mathcal{H}}

\newcommand{\cN}{\mathcal{N}}

\makeatletter
\gdef\@fpheader{}
\makeatother
\begin{document}
\title{Matter correlators through a wormhole in double-scaled SYK}

\author{Kazumi Okuyama}

\affiliation{Department of Physics, 
Shinshu University, 3-1-1 Asahi, Matsumoto 390-8621, Japan}

\emailAdd{kazumi@azusa.shinshu-u.ac.jp}

\abstract{We compute the two-point function of matter operators
in the double-scaled SYK (DSSYK) model, 
where the two matter operators
are inserted at each end of the cylindrical wormhole.
We find that the wormhole amplitude in DSSYK is written as a trace over the 
chord Hilbert space.
We also show that the length of the wormhole is stabilized in 
the semi-classical limit, by the same mechanism worked for 
the JT gravity case.
}

\maketitle

\section{Introduction}
Wormhole geometries have played an important role in the recent progress of
our understanding of quantum black holes and the Page curve
\cite{Penington:2019kki,Almheiri:2019qdq}.\footnote{See also
\cite{Stanford:2020wkf,deBoer:2023vsm,Sasieta:2022ksu,Cotler:2020lxj,Cotler:2021cqa,Cotler:2020hgz,Saad:2018bqo,Saad:2021rcu,Iliesiu:2021ari,Bah:2022uyz,Chen:2020ojn,Hsin:2020mfa,Gao:2016bin,Maldacena:2018lmt}, for example, 
for recent discussions on 
the role of wormhole geometries.}
However, in the AdS/CFT correspondence, the existence
of a wormhole connecting disjoint boundaries of bulk spacetime
leads to a puzzle, known as the factorization problem
\cite{Maldacena:2004rf}.
At least in lower dimensional examples of quantum gravity,
such as the JT gravity and the SYK model,
this factorization problem is naturally solved by assuming that
the boundary theory involves an ensemble average over the random
Hamiltonians. 
Indeed, as shown in \cite{Saad:2019lba} the JT gravity is holographycally
dual to a random matrix model,
where the random matrix plays the role of random Hamiltonians.

According to the classification in 
\cite{deBoer:2023vsm}, there are three types of averaging in the boundary theory:
(i) averaging over Hamiltonians, (ii) averaging over operators, and (iii) 
averaging over states.
In this paper, we consider the wormhole amplitude in the double-scaled SYK 
(DSSYK) model 
 \cite{Cotler:2016fpe,Berkooz:2018jqr}.
In this case, the wormhole 
 originates from the hybrid of averagings (i) and (ii) above.
We compute the two-point function of matter operators
in DSSYK, where the two matter operators
are inserted at each end of the cylindrical wormhole.
Our computation is a natural generalization of the similar
computation for JT gravity in \cite{Stanford:2020wkf}.
We find that the wormhole amplitude in DSSYK is written as a trace over the 
chord Hilbert space, which was interpreted in \cite{Lin:2022rbf}
as a Hilbert space of the bulk gravity theory.
Moreover, we find that the length of the wormhole is stabilized in 
the semi-classical limit, by the same mechanism worked for 
the JT gravity case \cite{Stanford:2020wkf}.

This paper is organized as follows. In section \ref{sec:review},
we review the result of DSSYK and its two-matrix model representation, 
known as the ETH matrix model.
In section \ref{sec:wormhole}, we compute the two-point function of matter operators
on the wormhole geometry.
In section \ref{sec:stabilization},
we study the saddle-point approximation of the wormhole amplitude and 
confirm that the length of the wormhole is stabilized at a non-zero length.
Finally we conclude in section \ref{sec:conclusion}
with some discussion on the future problems. 
In appendix \ref{app:Schwarzian} we consider 
the JT gravity limit of the wormhole amplitude.

\section{Review of DSSYK}\label{sec:review}
In this section we briefly review the result of DSSYK in \cite{Berkooz:2018jqr}
and the ETH matrix model for DSSYK in \cite{Jafferis:2022wez}.

The SYK model is defined by the Hamiltonian for 
$M$ Majorana fermions $\psi_i~(i=1,\cdots,M)$
obeying $\{\psi_i,\psi_j\}=2\cob_{i,j}$
with all-to-all $p$-body interaction
\begin{equation}
\begin{aligned}
 H=\ri^{p/2}\sum_{1\leq i_1<\cdots<i_p\leq M}
J_{i_1\cdots i_p}\psi_{i_1}\cdots\psi_{i_p},
\end{aligned} 
\end{equation}
where $J_{i_1\cdots i_p}$ is a random coupling drawn from the Gaussian distribution.
DSSYK is defined by the scaling limit
\begin{equation}
\begin{aligned}
 M,p\to\infty\quad\text{with}\quad \la=\frac{2p^2}{M}:\text{fixed}.
\end{aligned} 
\label{eq:scaling}
\end{equation}
As shown in \cite{Berkooz:2018jqr}, the ensemble average of the moment $\Tr H^k$ 
reduces to a counting problem
of the intersection number of chord diagrams
\begin{equation}
\begin{aligned}
 \bra \Tr H^k\ket_J=\sum_{\text{chord diagrams}}q^{\#(\text{intersections})}
\end{aligned} 
\end{equation}
with $q=e^{-\la}$.
This counting problem is solved by introducing the transfer matrix $T$
\begin{equation}
\begin{aligned}
 T=A_{+}+A_{-},
\end{aligned} 
\end{equation}
where $A_{\pm}$ denotes the 
$q$-deformed oscillator acting on the chord number state $|n\ket$
\begin{equation}
\begin{aligned}
 A_{+}|n\ket=\rt{\frac{1-q^{n+1}}{1-q}}|n+1\ket,\qquad
A_{-}|n\ket=\rt{\frac{1-q^{n}}{1-q}}|n-1\ket.
\end{aligned} 
\end{equation}
Then the disk amplitude of DSSYK is given by
\begin{equation}
\begin{aligned}
 \bra \Tr e^{-\bt H}\ket_J=\bra0|e^{\bt T}|0\ket.
\end{aligned} 
\label{eq:disk}
\end{equation}
The transfer matrix $T$ becomes diagonal in the $\th$-basis
\footnote{Our definition of $E(\th)$
in \eqref{eq:E-th} differs from that in \cite{Berkooz:2018jqr}
by the overall minus sign. Note that the sign of $E(\th)$ can
be flipped by
sending $\th\to\pi-\th$. Our definition \eqref{eq:E-th} is convenient to study
the low energy regime since the low energy limit
corresponds to $\th\to0$, 
not $\th\to\pi$.}
\begin{equation}
\begin{aligned}
 T|\th\ket=-E(\th)|\th\ket,\qquad E(\th)=-\frac{2\cos\th}{\rt{1-q}},
\end{aligned} 
\label{eq:E-th}
\end{equation}
and the overlap of $\bra n|$ and $|\th\ket$ is given by the $q$-Hermite polynomial
$H_n(\cos\th|q)$
\begin{equation}
\begin{aligned}
 \bra n|\th\ket=\frac{H_n(\cos\th|q)}{\rt{(q;q)_n}},
\end{aligned} 
\end{equation}
where the $q$-Pochhammer symbol is defined by
\begin{equation}
\begin{aligned}
 (a;q)_n=\prod_{k=0}^{n-1}(1-aq^k),\qquad
(a_1,\cdots,a_s;q)_n=\prod_{i=1}^s(a_i;q)_n.
\end{aligned} 
\end{equation}
$|\th\ket$ and $|n\ket$ are normalized as
\begin{equation}
\begin{aligned}
 \bra \th|\th'\ket=\frac{2\pi}{\mu(\th)}\cob(\th-\th'),\quad
\bra n|m\ket=\cob_{n,m},\quad
1=\int_0^\pi\frac{d\th}{2\pi}\mu(\th)|\th\ket\bra\th|=\sum_{n=0}^\infty
 |n\ket\bra n|,
\end{aligned} 
\label{eq:res-uni}
\end{equation}
and the measure factor $\mu(\th)$ is given by
\begin{equation}
\begin{aligned}
 \mu(\th)=(q,e^{\pm2\ri\th};q)_\infty.
\end{aligned} 
\label{eq:mu-th}
\end{equation}
Then the disk partition function \eqref{eq:disk} is written as
\begin{equation}
\begin{aligned}
 \bra0|e^{\bt T}|0\ket=\int_0^\pi\frac{d\th}{2\pi}\mu(\th)e^{-\bt E(\th)}.
\end{aligned} 
\end{equation}

As discussed in \cite{Berkooz:2018jqr}, we can also consider
the matter operator $\cO$
\begin{equation}
\begin{aligned}
 \cO=\ri^{s/2}\sum_{1\leq i_1<\cdots< i_s\leq M}K_{i_1\cdots i_s}\psi_{i_1}\cdots\psi_{i_s}
\end{aligned} 
\label{eq:matter-M}
\end{equation}
with Gaussian random coefficients $K_{i_1\cdots i_s}$
which is drawn independently from the random coupling
$J_{i_1\cdots i_p}$ in the SYK Hamiltonian.
In the double scaling limit \eqref{eq:scaling},
the effect of this operator can be made finite by taking the limit
$s\to\infty$ with $\lap=s/p$ held fixed.
Then the correlator of $\cO$'s is also written as a  
counting problem of the chord diagrams
\begin{equation}
\begin{aligned}
 \sum_{\text{chord diagrams}}q^{\#(H\text{-}H\,\text{intersections})}
q^{\lap\#(H\text{-}\cO\,\text{intersections})}
q^{\lap^2\#(\cO\text{-}\cO\,\text{intersections})}.
\end{aligned} 
\label{eq:chord-count}
\end{equation}
Note that there appear two types of chords in this computation: $H$-chords and 
$\cO$-chords coming from the Wick contraction of random couplings 
$J_{i_1\cdots i_p}$ and $K_{i_1\cdots i_s}$, respectively.
The $\cO$-chord is also called the matter chord.
For instance, the disk two-point function of $\cO$ is given by
\begin{equation}
\begin{aligned}
 \bra\Tr (e^{-\bt_1H}\cO e^{-\bt_2H}\cO)\ket_{J,K}=\bra 0|e^{\bt_1T}q^{\lap\h{N}}
e^{\bt_2T}|0\ket,
\end{aligned} 
\label{eq:disk-2pt}
\end{equation}
where $\h{N}$ is the number operator
\begin{equation}
\begin{aligned}
 \h{N}|n\ket=n|n\ket.
\end{aligned} 
\end{equation}
In terms of the $\th$-basis, the disk two-point function 
in \eqref{eq:disk-2pt}
is written as
\begin{equation}
\begin{aligned}
 \bra 0|e^{\bt_1T}q^{\lap\h{N}}
e^{\bt_2T}|0\ket=\int_0^\pi\prod_{i=1,2}\frac{d\th_i}{2\pi}\mu(\th_i)e^{-\bt_iE(\th_i)}
G(\th_1,\th_2)
\end{aligned} 
\label{eq:disk-2pt-G}
\end{equation}
with
\begin{equation}
\begin{aligned}
 G(\th_1,\th_2)=\bra\th_1|q^{\lap\h{N}}|\th_2\ket=\frac{(q^{2\lap};q)_\infty}
{(q^{\lap}e^{\ri(\pm\th_1\pm\th_2)};q)_\infty}.
\end{aligned}
\label{eq:G12-def} 
\end{equation}
In the rest of this paper, we will ignore
the intersections between the matter chords
for simplicity.
This amounts to consider a
bulk free field which is holographically dual to the matter operator $\cO$.

Next, let us consider the ETH matrix model for DSSYK introduced in \cite{Jafferis:2022wez}. As discussed in \cite{Jafferis:2022wez},
we can define the two-matrix model
\begin{equation}
\begin{aligned}
 \cZ=\int dAdB e^{- N\Tr V(A,B)}
\end{aligned} 
\label{eq:two-mat}
\end{equation}
which reproduces the disk partition function \eqref{eq:disk}
and the two-point function \eqref{eq:disk-2pt} of DSSYK in the large $N$ limit.
The $N\times N$ matrices $A$ and $B$ correspond to $H$ and $\cO$, respectively,
and the size $N$ of the matrices is given by the dimension of
the Fock space of $M$ Majorana fermions
\begin{equation}
\begin{aligned}
 N=2^{\frac{M}{2}}.
\end{aligned} 
\end{equation}
The potential $V(A,B)$ in \eqref{eq:two-mat}
is given by \cite{Jafferis:2022wez}
\begin{equation}
\begin{aligned}
 \Tr V(A,B)&=\sum_{a=1}^N\sum_{n=1}^\infty (-1)^{n-1}\left(1-\frac{q^{2n\lap}}{1-q^{2n}}\right)
\frac{q^{\hf n^2}}{n}(q^{\hf n}+q^{-\hf n})
T_{2n}\left(\frac{\rt{1-q}}{2}E_a\right)\\
&+\hf \sum_{a,b=1}^N B_{ab}B_{ba}G(\th_a,\th_b)^{-1},
\end{aligned} 
\label{eq:VAB}
\end{equation}
where $T_n(\cos\th)=\cos(n\th)$ denotes the Chebyshev polynomial of the first kind
and $E_a$ is the eigenvalue of the matrix $A$ given by $E(\th_a)$ in \eqref{eq:E-th}.
The first line of \eqref{eq:VAB}
is determined by requiring that the eigenvalue density of the matrix $A$
reproduces the measure factor $\mu(\th)$ in \eqref{eq:mu-th} in the large $N$
limit.
The second line of \eqref{eq:VAB} is constructed in such a way that
the two-point function of the matrix $B$
reproduces the disk two-point function of $\cO$ in \eqref{eq:disk-2pt-G}.

From \eqref{eq:VAB}, we can read off the propagator of the matrix $B$ as
\begin{equation}
\begin{aligned}
 \bra B_{ab}B_{cd}\ket=\frac{1}{N}G(\th_a,\th_b)\cob_{ad}\cob_{bc},
\end{aligned}
\label{eq:prop-B} 
\end{equation}
where the expectation value is taken in the two-matrix model
\begin{equation}
\begin{aligned}
 \bra\cdots\ket=\frac{1}{\cZ}\int dAdB(\cdots)e^{-N\Tr V(A,B)}.
\end{aligned} 
\end{equation}
We can check that the disk two-point function of $\cO$ in \eqref{eq:disk-2pt-G}
is reproduced from the two-matrix model
\begin{equation}
\begin{aligned}
 \big\bra \Tr(e^{-\bt_1A}Be^{-\bt_2A}B)\big\ket&=\sum_{a,b=1}^N\bra e^{-\bt_1E_a}B_{ab}
e^{-\bt_2E_b}B_{ba}\ket\\
&=\sum_{a,b=1}^N e^{-\bt_1E_a-\bt_2E_b}\frac{1}{N}G(\th_a,\th_b)\\
&=N\int_0^\pi\frac{d\th_1}{2\pi}\mu(\th_1)\int_0^\pi
\frac{d\th_2}{2\pi}\mu(\th_2)e^{-\bt_1E(\th_1)-\bt_2E(\th_2)}G(\th_1,\th_2).
\end{aligned} 
\label{eq:2pt-mat}
\end{equation}
Here we have used the relation
\begin{equation}
\begin{aligned}
 \sum_{a=1}^N f(E_a)=N\int_0^\pi\frac{d\th}{2\pi}\mu(\th)f(E(\th))
\end{aligned} 
\end{equation}
which is valid in the large $N$ limit.
In the next section, we compute the matter correlator of DSSYK on the wormhole geometry using the propagator in \eqref{eq:prop-B}.

\section{Matter correlators through a wormhole}\label{sec:wormhole}
Let us consider the matter correlator of DSSYK on the wormhole geometry.
In the two-matrix model description of DSSYK, it is given by 
$\big\bra \Tr (e^{-\bt_LA}B)\Tr (e^{-\bt_RA}B)\big\ket$.
Using the propagator of $B$ in \eqref{eq:prop-B} we find
\begin{equation}
\begin{aligned}
\big\bra \Tr (e^{-\bt_LA}B)\Tr (e^{-\bt_RA}B)\big\ket&=\sum_{a,b=1}^N
\bra e^{-\bt_LE_a}B_{aa} e^{-\bt_RE_b}B_{bb}\ket\\
&=\sum_{a=1}^Ne^{-(\bt_L+\bt_R)E_a}\frac{1}{N}G(\th_a,\th_a)\\
&=\int_0^\pi\frac{d\th}{2\pi}\mu(\th)e^{-(\bt_L+\bt_R) E(\th)}G(\th,\th).
\end{aligned} 
\label{eq:cyl-comp}
\end{equation}
Note that in this computation only the $a=b$ terms survive,
which is reminiscent of the Berry's ``diagonal approximation''
for the spectral form factor \cite{berry1985semiclassical}.
Note also that the power of $N$ in \eqref{eq:cyl-comp} is down by one
compared to the disk two-point function
in \eqref{eq:2pt-mat}.
Using \eqref{eq:res-uni} and \eqref{eq:G12-def}, we can easily
show that \eqref{eq:cyl-comp} is written as
\begin{equation}
\begin{aligned}
 \big\bra \Tr (e^{-\bt_LA}B)\Tr (e^{-\bt_RA}B)\big\ket
&=\Tr_{\cH}\bigl[e^{(\bt_L+\bt_R)T}q^{\lap\h{N}}\bigr],
\end{aligned} 
\label{eq:2pt-trace}
\end{equation}
where $\Tr_{\cH}$ denotes the trace over the Hilbert space $\cH$
spanned by the chord number states $\{|n\ket\}_{n=0,1,2\cdots}$
\begin{equation}
\begin{aligned}
 \cH=\bigoplus_{n=0}^\infty\mathbb{C}|n\ket.
\end{aligned} 
\end{equation}
More explicitly, the trace in \eqref{eq:2pt-trace} is written as
\begin{equation}
\begin{aligned}
 \Tr_{\cH}\bigl[e^{(\bt_L+\bt_R)T}q^{\lap\h{N}}\bigr]=\sum_{n=0}^\infty
q^{\lap n}\bra n|e^{(\bt_L+\bt_R)T}|n\ket.
\end{aligned} 
\label{eq:trace-sum}
\end{equation}
This amplitude
represents the cylindrical wormhole which is schematically depicted as
\begin{equation}
\begin{aligned}
 \begin{tikzpicture}[scale=0.8]
\draw (0,0) ellipse (0.5 and 1);
\draw (5,-1) arc (-90:90: 0.5 and 1);
\draw[dashed] (5,1) arc (90:270: 0.5 and 1);
\draw (0,1)--(5,1);
\draw (0,-1)--(5,-1);
\draw[thick,blue] (0.5,0)--(5.5,0);
\draw[blue,fill=blue] (0.5,0) circle (.3ex);
\draw[blue,fill=blue] (5.5,0) circle (.3ex);
\draw (0.5,0) node [left]{$\cO$};
\draw (5.5,0) node [right]{$\cO$};
\draw (3,0) node [above]{$n$};
\draw (5.3,0.9) node [right]{$\bt$};
\draw (-0.7,0) node [left]{$\displaystyle \sum_{n=0}^\infty$};
\end{tikzpicture}
\end{aligned}
\label{eq:cylinder}
\end{equation}
where the circumference $\bt$ of the cylinder is given by
\begin{equation}
\begin{aligned}
 \bt=\bt_L+\bt_R.
\end{aligned} 
\label{eq:bt-def}
\end{equation} 
In other words, the amplitude in \eqref{eq:2pt-trace}
represents the correlation of matter operators through the wormhole.
Our computation is a natural generalization of the similar computation in 
the JT gravity \cite{Stanford:2020wkf}.
In particular, the operator $q^{\lap\h{N}}$ in 
\eqref{eq:2pt-trace}
\begin{equation}
\begin{aligned}
 q^{\lap\h{N}}=\sum_{n=0}^\infty q^{\lap n}|n\ket\bra n|
\end{aligned} 
\end{equation}
is a discrete analogue of the operator $\h{G}_\cO$ appeared in 
\cite{Stanford:2020wkf} (see also \cite{Saad:2019pqd,Blommaert:2020seb})
\begin{equation}
\begin{aligned}
 \h{G}_\cO=\int_{-\infty}^\infty d\ell e^{-\lap\ell}|\ell\ket\bra\ell|,
\end{aligned} 
\end{equation}
where $\ell$ denotes the length of the bulk geodesic connecting the two operators $\cO$.
Note that $\ell$ can be negative since $\ell$ is a renormalized geodesic length.

We can insert more operators $\cO$
on the two boundaries of the cylinder
\begin{equation}
\begin{aligned}
 \Biggl\bra\Tr\prod_{i=1}^k e^{-\bt^L_{i}A}B
\Tr\prod_{i=1}^k e^{-\bt^R_{i}A}B\Biggr\ket_{\text{cylinder}}
=\sum_{\si\in S_k}
\Tr_\cH\left[\prod_{i=1}^k e^{(\bt^L_i+\bt^R_{\si(i)})T}q^{\lap \h{N}}\right].
\end{aligned}
\label{eq:k-corr} 
\end{equation}
For instance, when $k=2$ \eqref{eq:k-corr} becomes
\begin{equation}
\begin{aligned}
 &\Bigl\bra\Tr (e^{-\bt^L_1 A}B e^{-\bt^L_2A}B)\Tr(e^{-\bt^R_1 A}B e^{-\bt^R_2A}B)
\Bigr\ket_{\text{cylinder}}\\
=&\,\Tr_{\cH} \Bigl[e^{(\bt^L_1+\bt^R_1)T}q^{\lap \h{N}}
e^{(\bt^L_2+\bt^R_2)T}q^{\lap \h{N}}+
e^{(\bt^L_1+\bt^R_2)T}q^{\lap \h{N}}
e^{(\bt^L_2+\bt^R_1)T}q^{\lap \h{N}}\Bigr].
\end{aligned} 
\end{equation}

\section{Stabilization of the wormhole length}\label{sec:stabilization}
In general, the wormhole geometry with a fixed length 
is not a classical solution of the bulk gravity theory. 
However, as discussed in \cite{Stanford:2020wkf}
for the JT gravity case, the length of the wormhole 
is stabilized by including the effect of matter correlator.
This is similar in spirit to the 
construction of wormhole solutions as  
gravitational constrained instantons
\cite{Cotler:2020lxj,Cotler:2021cqa}.
Below we will see that the length $n$ of the wormhole 
in \eqref{eq:trace-sum} is stabilized in the semi-classical 
saddle-point approximation.

The wormhole amplitude in \eqref{eq:trace-sum}
is written as
\begin{equation}
\begin{aligned}
 \Tr_\cH\bigl(e^{\bt T}q^{\lap\h{N}}\bigr)
=\sum_{n=0}^\infty q^{\lap n}\int_0^\pi\frac{d\th}{2\pi}\mu(\th)|\bra n|\th\ket|^2e^{-\bt E(\th)},
\end{aligned} 
\label{eq:2pt-th-int}
\end{equation}
where $\bt$ is defined in \eqref{eq:bt-def}.
Let us consider the saddle-point approximation of the $\th$-integral
in the semi-classical limit $\la\to0$.
Using the relation
\begin{equation}
\begin{aligned}
 (a;q)_\infty=\exp\left(-\sum_{n=1}^\infty\frac{a^n}{n(1-q^n)}\right)
\approx e^{-\frac{1}{\la}\text{Li}_2(a)},
\end{aligned} 
\end{equation}
the measure factor $\mu(\th)$ is approximated as
\begin{equation}
\begin{aligned}
 \mu(\th)&=(q,e^{\pm2\ri\th};q)_\infty\approx 
\exp\left[-\frac{1}{\la}\sum_{\vep=\pm1}\text{Li}_2(e^{2\vep\ri\th})\right].
\end{aligned} 
\end{equation}
Here $\text{Li}_2(z)$ 
denotes the dilogarithm function.
Also, using the relation
\begin{equation}
\begin{aligned}
  \sum_{n=0}^\infty t^n |\bra n|\th\ket|^2
=\frac{(t^2;q)_\infty}{(te^{\pm 2\ri\th};q)_\infty (t;q)_\infty^2},
\end{aligned} 
\end{equation}
the factor $|\bra n|\th\ket|^2$ in 
\eqref{eq:2pt-th-int}
is written as
\begin{equation}
\begin{aligned}
 |\bra n|\th\ket|^2
&=\int_0^{2\pi}\frac{d\phi}{2\pi}e^{-\ri n\phi}
\frac{(e^{2\ri\phi};q)_\infty}{(e^{\ri\phi\pm 2\ri\th};q)_\infty 
(e^{\ri\phi};q)_\infty^2}\\
&\approx
\int_0^{2\pi}\frac{d\phi}{2\pi}e^{-\ri n\phi}
\exp\left[
-\frac{1}{\la}\text{Li}_2(e^{2\ri\phi})
+\frac{1}{\la}\sum_{\vep=\pm1}\text{Li}_2(e^{\ri\phi+2\vep\ri\th})
+\frac{2}{\la}\text{Li}_2(e^{\ri\phi})\right].
\end{aligned} 
\end{equation}
If we further assume that
\begin{equation}
\begin{aligned}
 n\sim\cO(\la^{-1}),\quad \bt \sim\cO(\la^0),
\end{aligned} 
\label{eq:scale-nbt}
\end{equation}
the integral in \eqref{eq:2pt-th-int} is written as
\begin{equation}
\begin{aligned}
\int_0^\pi \frac{d\th}{2\pi}\mu(\th)|\bra n|\th\ket|^2e^{-\bt E(\th)}
\approx 
\int_0^\pi \frac{d\th}{2\pi}
\int_0^{2\pi} \frac{d\phi}{2\pi}e^{-\frac{1}{\la}F(\th,\phi)-\bt E(\th)},
\end{aligned} 
\label{eq:F-int}
\end{equation}
where
\begin{equation}
\begin{aligned}
 F(\th,\phi)=\ri\la n\phi+
\text{Li}_2(e^{2\ri\phi})-2\text{Li}_2(e^{\ri\phi})+\sum_{\vep=\pm1}
\Bigl[\text{Li}_2(e^{2\vep\ri\th})-\text{Li}_2(e^{\ri\phi+2\vep\ri\th})\Bigr].
\end{aligned} 
\end{equation}
We can easily show that the solution of the saddle-point equation
\begin{equation}
\begin{aligned}
 \del_\th F(\th,\phi)\Big|_{\th=\th_*,\phi=\phi_*}=\del_\phi F(\th,\phi)
\Big|_{\th=\th_*,\phi=\phi_*}=0
\end{aligned} 
\label{eq:F-saddle}
\end{equation}
is given by
\begin{equation}
\begin{aligned}
 \phi_*=0,\quad \sin^2\th_*=q^n.
\end{aligned} 
\label{eq:saddle-thphi}
\end{equation}
Then $E(\th)$ in \eqref{eq:2pt-th-int}
is approximated as
\begin{equation}
\begin{aligned}
 E(\th_*)=-\frac{2\cos\th_*}{\rt{1-q}}=-2\rt{\frac{1-q^n}{1-q}}.
\end{aligned} 
\label{eq:E-th*}
\end{equation}
Alternatively, the same approximation can be obtained from the recursion relation
obeyed by $\psi_n(\th)=\bra n|\th\ket$
\begin{equation}
\begin{aligned}
 2\cos\th\psi_{n}(\th)=\rt{1-q^n}\psi_{n-1}(\th)
+\rt{1-q^{n+1}}\psi_{n+1}(\th).
\end{aligned} 
\label{eq:rec-psin} 
\end{equation}
From this relation, $\cos\th$ is approximated as
\begin{equation}
\begin{aligned}
 \cos\th\approx\frac{\rt{1-q^n}+\rt{1-q^{n+1}}}{2}\approx\rt{1-q^n},
\end{aligned} 
\end{equation} 
which agrees with the result of saddle-point approximation 
\eqref{eq:saddle-thphi}.
We should stress that our saddle-point analysis is different from 
\cite{Goel:2018ubv}.
In \cite{Goel:2018ubv}, the $\bt E(\th)$ term is assumed to be of order 
$\cO(\la^{-1})$ and hence the saddle-point value of $\th$ depends on $\bt$.
In our case we assumed the scaling in \eqref{eq:scale-nbt},
which implies that $\bt E(\th)$ is sub-leading compared to the leading
term $\la^{-1}F(\th,\phi)$
in \eqref{eq:F-int}.
Thus the $\bt E(\th)$ term does not enter into the saddle-point equation 
\eqref{eq:F-saddle}.

\begin{figure}[htb]
\centering
\includegraphics
[width=0.5\linewidth]{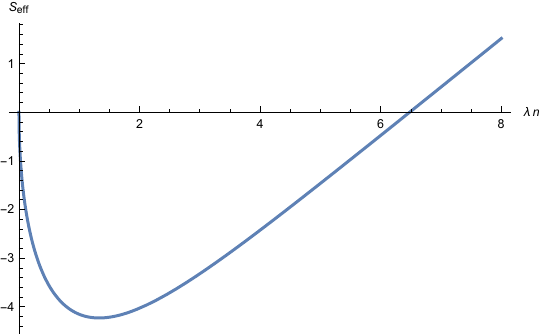}
  \caption{
Plot of $S_{\text{eff}}(n)$ in \eqref{eq:Seff-n} as a function of $\la n$.
In this figure we have set $\lap=1,\bt=1,\la=1/10$.
}
  \label{fig:Seff}
\end{figure}

Finally, we arrive at the saddle-point approximation
of the wormhole amplitude
\begin{equation}
\begin{aligned}
 \Tr_\cH\bigl(e^{\bt T}q^{\lap\h{N}}\bigr)
\approx\sum_{n=0}^\infty q^{\lap n}e^{-\bt E(\th_*)}
\equiv \sum_{n=0}^\infty e^{-S_{\text{eff}}(n)},
\end{aligned} 
\label{eq:sum-approx}
\end{equation}
where
\begin{equation}
\begin{aligned}
 S_{\text{eff}}(n)
&=\lap\la n-2\bt\rt{\frac{1-e^{-\la n}}{1-e^{-\la}}}.
\end{aligned} 
\label{eq:Seff-n}
\end{equation}
In Figure \ref{fig:Seff}, we show the plot of $S_{\text{eff}}(n)$ in \eqref{eq:Seff-n}.
We can see that $S_{\text{eff}}(n)$ has a minimum at a non-zero value of $\la n$
and hence the wormhole length $n$ is stabilized 
by the effect of matter correlator.
This stabilization mechanism of the wormhole length for DSSYK is 
essentially the same as that for the JT gravity discussed in \cite{Stanford:2020wkf}
(see appendix \ref{app:Schwarzian} for the JT gravity limit of the wormhole amplitude
\eqref{eq:2pt-th-int}).

From \eqref{eq:Seff-n},
we find that the solution $n=n_*$ of the saddle-point equation
$\del_nS_{\text{eff}}(n)=0$ is given by
\begin{equation}
\begin{aligned}
 q^{n_*}=\frac{2\lap}{\lap+\rt{\lap^2+\tilde{\bt}^2}},
\end{aligned} 
\label{eq:n*}
\end{equation}
where $\tilde{\bt}$ is defined by
\begin{equation}
\begin{aligned}
 \tilde{\bt}=\frac{2\bt}{\rt{1-q}}.
\end{aligned} 
\end{equation}
We can also estimate the variance of the wormhole length
from the second derivative of $S_{\text{eff}}(n)$ at $n=n_*$
\begin{equation}
\begin{aligned}
 \bra(n-n_*)^2\ket\sim\frac{1}{\del_n^2S_{\text{eff}}(n_*)}
=\frac{1}{\la^2}\left[\lap+\frac{\lap^2}{\tilde{\bt}^2}
\bigl(\lap+\rt{\lap^2+\tilde{\bt}^2}\bigr)\right]^{-1}.
\end{aligned} 
\end{equation}

As discussed in \cite{Stanford:2020wkf}, 
the $\bt\to0$ limit of the wormhole amplitude in JT gravity 
is UV divergent since the cylinder becomes very thin as $\bt\to0$. 
Interestingly, in the case of DSSYK the wormhole amplitude in 
\eqref{eq:2pt-trace} is finite 
in this limit
\begin{equation}
\begin{aligned}
 \lim_{\bt\to0}\Tr_{\cH}\bigl(e^{\bt T}q^{\lap\h{N}}\bigr)
=\Tr_\cH \bigl(q^{\lap\h{N}}\bigr)=\sum_{n=0}^\infty q^{\lap n}=\frac{1}{1-q^\lap}.
\label{eq:thin-lim}
\end{aligned} 
\end{equation}
Thus, DSSYK can be thought of as a UV completion of the JT gravity.
As discussed in \cite{Lin:2022rbf,Jafferis:2022wez,Berkooz:2022mfk,Okuyama:2023byh},
the UV finiteness of DSSYK is closely related to the existence of
a minimal length in the bulk geometry.
Note that the wormhole solution with a finite length ceases to exist
in the thin cylinder limit $\bt\to0$
since the saddle-point solution in \eqref{eq:n*} 
collapses to a zero-length wormhole $n_*=0$.
In other words, 
the semiclassical approximation breaks down in this limit. Instead,
we have to sum over all $n$ as shown in \eqref{eq:thin-lim}.

\section{Conclusion and outlook}\label{sec:conclusion}
In this paper we have studied the matter two-point function through
the wormhole geometry in DSSYK. We find that this two-point function 
\eqref{eq:2pt-trace} is 
written as a trace on the Hilbert space spanned by the chord number states,
which can be identified as the Hilbert space of the bulk gravity theory 
\cite{Lin:2022rbf}. In the saddle-point approximation,
the sum over $n$ in \eqref{eq:trace-sum}
is dominated by a non-zero value of $n$
due to the balance between the matter contribution $q^{\lap n}$
and the Boltzmann factor $e^{-\bt E(\th_*)}$ in \eqref{eq:sum-approx}.
This stabilization mechanism of the wormhole length is the same as that appeared in 
the computation of the wormhole amplitude in JT gravity \cite{Stanford:2020wkf}.
This is similar in spirit to the gravitational constrained instantons in 
\cite{Cotler:2020lxj,Cotler:2021cqa}.
One important difference from \cite{Stanford:2020wkf} is that
the wormhole amplitude \eqref{eq:2pt-trace}
in DSSYK is finite even in the thin cylinder limit \eqref{eq:thin-lim}.

There are many interesting open questions.
In this paper, we have ignored the intersections of matter chords, 
which amounts to consider the free field in the bulk theory.
It would be interesting to incorporate the matter interaction and see how the result in this paper is modified.
It would also be 
interesting to generalize our computation to higher genus topologies
along the lines of \cite{Okuyama:2023kdo,Okuyama:2023aup}.
The higher dimensional generalization 
of our computation would be also interesting (see e.g. \cite{Cotler:2020hgz}
for a wormhole in the AdS$_3$/CFT$_{2}$ correspondence).

We have seen that the stabilization mechanism  of the wormhole length advocated in 
\cite{Stanford:2020wkf} works for DSSYK as well. However, we should stress
that the existence of a semi-classical saddle in the gravitational path integral
is not a necessary condition for the definition of quantum gravity.
In our case, the wormhole length $n$ is a dynamical variable and the wormhole
amplitude \eqref{eq:trace-sum} is defined by summing over all possible $n$.
The wormhole
amplitude \eqref{eq:trace-sum} itself is well-defined without
requiring the existence of a semi-classical saddle.
In fact, the wormhole
amplitude of DSSYK is finite
in the thin cylinder limit \eqref{eq:thin-lim},
which is far from the semi-classical regime.
By definition, quantum gravity should make sense in the quantum regime, which
is indeed the case in our computation of 
the wormhole amplitude in DSSYK.
\acknowledgments
This work was supported
in part by JSPS Grant-in-Aid for Transformative Research Areas (A) 
``Extreme Universe'' 21H05187 and JSPS KAKENHI Grant 22K03594.

\appendix
\section{JT gravity limit}\label{app:Schwarzian}
In this appendix, we consider the JT gravity limit of the wormhole amplitude
in \eqref{eq:2pt-th-int}. The wavefunction
$\psi_n(\th)=\bra n|\th\ket$
obeys the recursion relation \eqref{eq:rec-psin}.
If we set
\begin{equation}
\begin{aligned}
 \th=\la s,\quad q^n=\la^2 e^{-\ell},
\end{aligned} 
\label{eq:s-ell}
\end{equation} 
and take the limit $\la\to0$ with $s$ and $\ell$ held fixed, 
then the $\cO(\la^2)$ term of \eqref{eq:rec-psin} becomes
\begin{equation}
\begin{aligned}
 s^2\psi=\bigl(-\del_\ell^2+e^{-\ell}\bigr)\psi.
\end{aligned} 
\end{equation} 
This is solved by the modified Bessel function
\begin{equation}
\begin{aligned}
 \psi=K_{2\ri s}(2e^{-\hf\ell}),
\end{aligned} 
\end{equation}
which reproduces the known result of the wavefunction in JT gravity.

The $\la\to0$ limit of $\mu(\th)$ in \eqref{eq:mu-th}
can be obtained by rewriting it as
\begin{equation}
\begin{aligned}
 \mu(\th)=2\sin\th q^{-\frac{1}{8}}
\vth_1\left(\frac{\th}{\pi},\frac{\ri\la}{2\pi}\right),
\end{aligned} 
\end{equation}
where $\vth_1(z,\tau)$ is the Jacobi theta function
\begin{equation}
\begin{aligned}
 \vth_1(z,\tau)=2q^{\frac{1}{8}}\sin\pi z \prod_{n=1}^\infty
(1-q^n)(1-e^{2\pi\ri z}q^n)(1-e^{-2\pi\ri z}q^n).
\end{aligned} 
\end{equation}
Using the transformation of Jacobi theta function
\begin{equation}
\begin{aligned}
 \vth_1(z,\tau)=\ri (-\ri\tau)^{-\hf}e^{-\frac{\pi\ri z^2}{\tau}}
\vth_1\left(\frac{z}{\tau},-\frac{1}{\tau}\right),
\end{aligned} 
\end{equation}
the measure factor $\mu(\th)$ becomes
\begin{equation}
\begin{aligned}
 \mu(\th)=4\sin (\la s)\sinh(2\pi s)
\rt{\frac{2\pi}{\la}}e^{-2\la s^2+\frac{\la}{8}-\frac{\pi^2}{2\la}}
\prod_{n=1}^\infty (1-e^{-\frac{4\pi^2 n}{\la}})
(1-e^{4\pi s-\frac{4\pi^2 n}{\la}})
(1-e^{-4\pi s-\frac{4\pi^2 n}{\la}}),
\end{aligned} 
\label{eq:mu-Jacobi}
\end{equation}
where $\th$ and $s$ are related by \eqref{eq:s-ell}.
Finally, taking the $\la\to0$ limit of \eqref{eq:mu-Jacobi}
we find
\begin{equation}
\begin{aligned}
 \lim_{\la\to0}\mu(\th)=\cN\cdot s\sinh(2\pi s),
\end{aligned} 
\end{equation}
where $\cN=4\rt{2\pi\la}e^{-\frac{\pi^2}{2\la}}$ is the
overall normalization coefficient.
Then the $\la\to0$ limit of the wormhole amplitude \eqref{eq:2pt-th-int} becomes
\begin{equation}
\begin{aligned}
 \big\bra\Tr (e^{-\bt_L H}\cO)\Tr(e^{-\bt_R H}\cO)\big\ket
=\int_{-\infty}^\infty d\ell e^{-\lap \ell}
\int_0^\infty ds s\sinh(2\pi s)\big|K_{2\ri s}(2e^{-\hf\ell})\big|^2 e^{-\bt' s^2},
\end{aligned} 
\label{eq:cyl-sint}
\end{equation}
where $\bt'=\bt\la^{\frac{3}{2}}$ and we ignored the overall normalization. 
Note that $K_{2\ri s}(2e^{-\hf\ell})$ is invariant under $s\to-s$
\begin{equation}
\begin{aligned}
 K_{-2\ri s}(2e^{-\hf\ell})=K_{2\ri s}(2e^{-\hf\ell}).
\end{aligned} 
\end{equation}

\begin{figure}[htb]
\centering
\includegraphics
[width=0.5\linewidth]{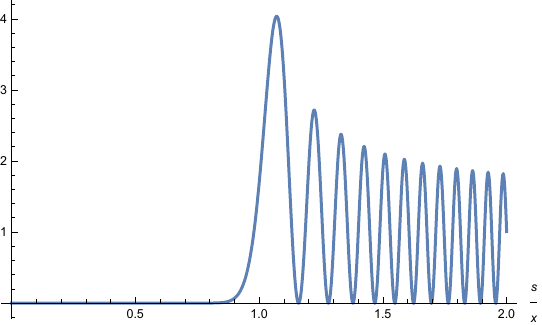}
  \caption{
Plot of $s\sinh(2\pi s)|K_{2\ri s}(2x)|^2$ as a function of $s$ for a fixed $x$.
The horizontal axis is $s/x$.
In this figure we set $x=20$. 
}
  \label{fig:bessel}
\end{figure}

Now, let us consider the saddle-point approximation of the
$s$-integral in \eqref{eq:cyl-sint}.
To this end, we need the asymptotic form of $K_{2\ri s}(2x)$ in the regime
where
\begin{equation}
\begin{aligned}
 s,x\gg1,\quad \frac{s}{x}=\text{fixed}.
\end{aligned} 
\end{equation}
Using the integral representation of the modified Bessel function
\begin{equation}
\begin{aligned}
 K_{2\ri s}(2x)=\hf\int_{-\infty}^\infty dt \,e^{2\ri st-2x\coth t},
\end{aligned} 
\label{eq:K-int}
\end{equation}
we find that $K_{2\ri s}(2x)$ behaves differently for $s<x$ and $s>x$.
When $s<x$, the $t$-integral
in \eqref{eq:K-int} has a saddle-point at
\begin{equation}
\begin{aligned}
t_*=\ri u,\quad \sin u=\frac{s}{x}.
\end{aligned} 
\label{eq:saddle-u}
\end{equation}
When $s>x$, the saddle-point is given by
\begin{equation}
\begin{aligned}
 t_*=\frac{\pi\ri}{2}+v,\quad \cosh v=\frac{s}{x}. 
\end{aligned} 
\label{eq:saddle-v}
\end{equation}
From \eqref{eq:saddle-u} and \eqref{eq:saddle-v}, we find that
$K_{2\ri s}(2x)$ behaves as \footnote{One can check that the
large $s$ expansion of the second line of \eqref{eq:K-asy}
is consistent with the asymptotic formula of the modified Bessel function in 
\cite{wolfram}.}
\begin{equation}
K_{2\ri s}(2x)\sim\left\{
\begin{aligned}
 &e^{-2s(u+\cot u)} ,\quad &(s<x),\\
& \text{Re}\bigl[e^{-\pi s+2\ri s(v-\tanh v)}\bigr],\quad &(s>x).
\end{aligned} \right.
\label{eq:K-asy}
\end{equation}
It turns out that $s\sinh(2\pi s)|K_{2\ri s}(2x)|^2$
has a maximum around $s= x$ as a function of $s$ for a fixed $x$
(see Figure \ref{fig:bessel}). 
Thus  the $s$-integral in \eqref{eq:cyl-sint} is dominated by
the saddle-point at
\begin{equation}
\begin{aligned}
 E(s_*)=s_*^2= e^{-\ell}.
\end{aligned} 
\label{eq:s*}
\end{equation}
This agrees with the claim in \cite{Stanford:2020wkf}.\footnote{
This result \eqref{eq:s*} was briefly mentioned in \cite{Stanford:2020wkf}
without a detailed derivation.}
The same result can be obtained by taking the $\la\to0$ limit of $\sin^2\th_*$
in \eqref{eq:saddle-thphi}
with $s,\ell$ in \eqref{eq:s-ell} held fixed.

\bibliography{paper}

\providecommand{\href}[2]{#2}\begingroup\raggedright\begin{thebibliography}{10}

\bibitem{Penington:2019kki}
G.~Penington, S.~H. Shenker, D.~Stanford, and Z.~Yang, ``{Replica wormholes and
  the black hole interior},''
  \href{http://dx.doi.org/10.1007/JHEP03(2022)205}{{\em JHEP} {\bfseries 03}
  (2022) 205}, \href{http://arxiv.org/abs/1911.11977}{{\ttfamily
  arXiv:1911.11977 [hep-th]}}.

\bibitem{Almheiri:2019qdq}
A.~Almheiri, T.~Hartman, J.~Maldacena, E.~Shaghoulian, and A.~Tajdini,
  ``{Replica Wormholes and the Entropy of Hawking Radiation},''
  \href{http://dx.doi.org/10.1007/JHEP05(2020)013}{{\em JHEP} {\bfseries 05}
  (2020) 013}, \href{http://arxiv.org/abs/1911.12333}{{\ttfamily
  arXiv:1911.12333 [hep-th]}}.

\bibitem{Stanford:2020wkf}
D.~Stanford, ``{More quantum noise from wormholes},''
  \href{http://arxiv.org/abs/2008.08570}{{\ttfamily arXiv:2008.08570
  [hep-th]}}.

\bibitem{deBoer:2023vsm}
J.~de~Boer, D.~Liska, B.~Post, and M.~Sasieta, ``{A principle of maximum
  ignorance for semiclassical gravity},''
  \href{http://arxiv.org/abs/2311.08132}{{\ttfamily arXiv:2311.08132
  [hep-th]}}.

\bibitem{Sasieta:2022ksu}
M.~Sasieta, ``{Wormholes from heavy operator statistics in AdS/CFT},''
  \href{http://dx.doi.org/10.1007/JHEP03(2023)158}{{\em JHEP} {\bfseries 03}
  (2023) 158}, \href{http://arxiv.org/abs/2211.11794}{{\ttfamily
  arXiv:2211.11794 [hep-th]}}.

\bibitem{Cotler:2020lxj}
J.~Cotler and K.~Jensen, ``{Gravitational Constrained Instantons},''
  \href{http://dx.doi.org/10.1103/PhysRevD.104.L081501}{{\em Phys. Rev. D}
  {\bfseries 104} (2021) 081501},
  \href{http://arxiv.org/abs/2010.02241}{{\ttfamily arXiv:2010.02241
  [hep-th]}}.

\bibitem{Cotler:2021cqa}
J.~Cotler and K.~Jensen, ``{Wormholes and black hole microstates in AdS/CFT},''
  \href{http://dx.doi.org/10.1007/JHEP09(2021)001}{{\em JHEP} {\bfseries 09}
  (2021) 001}, \href{http://arxiv.org/abs/2104.00601}{{\ttfamily
  arXiv:2104.00601 [hep-th]}}.

\bibitem{Cotler:2020hgz}
J.~Cotler and K.~Jensen, ``{AdS$_3$ wormholes from a modular bootstrap},''
  \href{http://dx.doi.org/10.1007/JHEP11(2020)058}{{\em JHEP} {\bfseries 11}
  (2020) 058}, \href{http://arxiv.org/abs/2007.15653}{{\ttfamily
  arXiv:2007.15653 [hep-th]}}.

\bibitem{Saad:2018bqo}
P.~Saad, S.~H. Shenker, and D.~Stanford, ``{A semiclassical ramp in SYK and in
  gravity},'' \href{http://arxiv.org/abs/1806.06840}{{\ttfamily
  arXiv:1806.06840 [hep-th]}}.

\bibitem{Saad:2021rcu}
P.~Saad, S.~H. Shenker, D.~Stanford, and S.~Yao, ``{Wormholes without
  averaging},'' \href{http://arxiv.org/abs/2103.16754}{{\ttfamily
  arXiv:2103.16754 [hep-th]}}.

\bibitem{Iliesiu:2021ari}
L.~V. Iliesiu, M.~Mezei, and G.~S\'arosi, ``{The volume of the black hole
  interior at late times},''
  \href{http://dx.doi.org/10.1007/JHEP07(2022)073}{{\em JHEP} {\bfseries 07}
  (2022) 073}, \href{http://arxiv.org/abs/2107.06286}{{\ttfamily
  arXiv:2107.06286 [hep-th]}}.

\bibitem{Bah:2022uyz}
I.~Bah, Y.~Chen, and J.~Maldacena, ``{Estimating global charge violating
  amplitudes from wormholes},''
  \href{http://dx.doi.org/10.1007/JHEP04(2023)061}{{\em JHEP} {\bfseries 04}
  (2023) 061}, \href{http://arxiv.org/abs/2212.08668}{{\ttfamily
  arXiv:2212.08668 [hep-th]}}.

\bibitem{Chen:2020ojn}
Y.~Chen and H.~W. Lin, ``{Signatures of global symmetry violation in relative
  entropies and replica wormholes},''
  \href{http://dx.doi.org/10.1007/JHEP03(2021)040}{{\em JHEP} {\bfseries 03}
  (2021) 040}, \href{http://arxiv.org/abs/2011.06005}{{\ttfamily
  arXiv:2011.06005 [hep-th]}}.

\bibitem{Hsin:2020mfa}
P.-S. Hsin, L.~V. Iliesiu, and Z.~Yang, ``{A violation of global symmetries
  from replica wormholes and the fate of black hole remnants},''
  \href{http://dx.doi.org/10.1088/1361-6382/ac2134}{{\em Class. Quant. Grav.}
  {\bfseries 38} no.~19, (2021) 194004},
  \href{http://arxiv.org/abs/2011.09444}{{\ttfamily arXiv:2011.09444
  [hep-th]}}.

\bibitem{Gao:2016bin}
P.~Gao, D.~L. Jafferis, and A.~C. Wall, ``{Traversable Wormholes via a Double
  Trace Deformation},'' \href{http://dx.doi.org/10.1007/JHEP12(2017)151}{{\em
  JHEP} {\bfseries 12} (2017) 151},
  \href{http://arxiv.org/abs/1608.05687}{{\ttfamily arXiv:1608.05687
  [hep-th]}}.

\bibitem{Maldacena:2018lmt}
J.~Maldacena and X.-L. Qi, ``{Eternal traversable wormhole},''
  \href{http://arxiv.org/abs/1804.00491}{{\ttfamily arXiv:1804.00491
  [hep-th]}}.

\bibitem{Maldacena:2004rf}
J.~M. Maldacena and L.~Maoz, ``{Wormholes in AdS},''
  \href{http://dx.doi.org/10.1088/1126-6708/2004/02/053}{{\em JHEP} {\bfseries
  02} (2004) 053}, \href{http://arxiv.org/abs/hep-th/0401024}{{\ttfamily
  arXiv:hep-th/0401024}}.

\bibitem{Saad:2019lba}
P.~Saad, S.~H. Shenker, and D.~Stanford, ``{JT gravity as a matrix integral},''
  \href{http://arxiv.org/abs/1903.11115}{{\ttfamily arXiv:1903.11115
  [hep-th]}}.

\bibitem{Cotler:2016fpe}
J.~S. Cotler, G.~Gur-Ari, M.~Hanada, J.~Polchinski, P.~Saad, S.~H. Shenker,
  D.~Stanford, A.~Streicher, and M.~Tezuka, ``{Black Holes and Random
  Matrices},'' \href{http://dx.doi.org/10.1007/JHEP05(2017)118}{{\em JHEP}
  {\bfseries 05} (2017) 118}, \href{http://arxiv.org/abs/1611.04650}{{\ttfamily
  arXiv:1611.04650 [hep-th]}}. [Erratum: JHEP 09, 002 (2018)].

\bibitem{Berkooz:2018jqr}
M.~Berkooz, M.~Isachenkov, V.~Narovlansky, and G.~Torrents, ``{Towards a full
  solution of the large N double-scaled SYK model},''
  \href{http://dx.doi.org/10.1007/JHEP03(2019)079}{{\em JHEP} {\bfseries 03}
  (2019) 079}, \href{http://arxiv.org/abs/1811.02584}{{\ttfamily
  arXiv:1811.02584 [hep-th]}}.

\bibitem{Lin:2022rbf}
H.~W. Lin, ``{The bulk Hilbert space of double scaled SYK},''
  \href{http://dx.doi.org/10.1007/JHEP11(2022)060}{{\em JHEP} {\bfseries 11}
  (2022) 060}, \href{http://arxiv.org/abs/2208.07032}{{\ttfamily
  arXiv:2208.07032 [hep-th]}}.

\bibitem{Jafferis:2022wez}
D.~L. Jafferis, D.~K. Kolchmeyer, B.~Mukhametzhanov, and J.~Sonner, ``{JT
  gravity with matter, generalized ETH, and Random Matrices},''
  \href{http://arxiv.org/abs/2209.02131}{{\ttfamily arXiv:2209.02131
  [hep-th]}}.

\bibitem{berry1985semiclassical}
M.~V. Berry, ``Semiclassical theory of spectral rigidity,''
  \href{http://dx.doi.org/10.1098/rspa.1985.0078}{{\em Proceedings of the Royal
  Society of London. A. Mathematical and Physical Sciences} {\bfseries 400}
  no.~1819, (1985) 229--251}.

\bibitem{Saad:2019pqd}
P.~Saad, ``{Late Time Correlation Functions, Baby Universes, and ETH in JT
  Gravity},'' \href{http://arxiv.org/abs/1910.10311}{{\ttfamily
  arXiv:1910.10311 [hep-th]}}.

\bibitem{Blommaert:2020seb}
A.~Blommaert, ``{Dissecting the ensemble in JT gravity},''
  \href{http://dx.doi.org/10.1007/JHEP09(2022)075}{{\em JHEP} {\bfseries 09}
  (2022) 075}, \href{http://arxiv.org/abs/2006.13971}{{\ttfamily
  arXiv:2006.13971 [hep-th]}}.

\bibitem{Goel:2018ubv}
A.~Goel, H.~T. Lam, G.~J. Turiaci, and H.~Verlinde, ``{Expanding the Black Hole
  Interior: Partially Entangled Thermal States in SYK},''
  \href{http://dx.doi.org/10.1007/JHEP02(2019)156}{{\em JHEP} {\bfseries 02}
  (2019) 156}, \href{http://arxiv.org/abs/1807.03916}{{\ttfamily
  arXiv:1807.03916 [hep-th]}}.

\bibitem{Berkooz:2022mfk}
M.~Berkooz, M.~Isachenkov, P.~Narayan, and V.~Narovlansky, ``{Quantum groups,
  non-commutative $AdS_2$, and chords in the double-scaled SYK model},''
  \href{http://arxiv.org/abs/2212.13668}{{\ttfamily arXiv:2212.13668
  [hep-th]}}.

\bibitem{Okuyama:2023byh}
K.~Okuyama, ``{End of the world brane in double scaled SYK},''
  \href{http://dx.doi.org/10.1007/JHEP08(2023)053}{{\em JHEP} {\bfseries 08}
  (2023) 053}, \href{http://arxiv.org/abs/2305.12674}{{\ttfamily
  arXiv:2305.12674 [hep-th]}}.

\bibitem{Okuyama:2023kdo}
K.~Okuyama, ``{Discrete analogue of the Weil-Petersson volume in double scaled
  SYK},'' \href{http://dx.doi.org/10.1007/JHEP09(2023)133}{{\em JHEP}
  {\bfseries 09} (2023) 133}, \href{http://arxiv.org/abs/2306.15981}{{\ttfamily
  arXiv:2306.15981 [hep-th]}}.

\bibitem{Okuyama:2023aup}
K.~Okuyama and T.~Suyama, ``{Solvable limit of ETH matrix model for
  double-scaled SYK},'' \href{http://arxiv.org/abs/2311.02846}{{\ttfamily
  arXiv:2311.02846 [hep-th]}}.

\bibitem{wolfram}
``{The Wolfram Functions Site: Bessel-Type Functions}.''
  \url{https://functions.wolfram.com/Bessel-TypeFunctions/BesselK/06/02/02/}.

\end{thebibliography}\endgroup
\bibliographystyle{utphys}

\end{document}